\documentclass[prl, 10pt,aps,nofootinbib, floatfix, notitlepage,twocolumn,longbibliography,superscriptaddress]{revtex4-2}

\usepackage{amsmath,amsfonts,amssymb}
\usepackage[colorlinks=true,citecolor=blue,urlcolor=blue]{hyperref}
\usepackage{graphicx}
\usepackage{mathrsfs}
\usepackage{xcolor}
\usepackage{mathtools}
\usepackage{orcidlink}
\usepackage[utf8]{inputenc}
\usepackage{booktabs}
\usepackage{microtype}

\newcommand{\bB}{{\bf B}}

\newcommand{\bE}{{\bf E}}

\long\def\exclude#1{}

\newcommand{\Eav}{E_{\rm av}}

\makeatletter 
    
\renewcommand\onecolumngrid{
\do@columngrid{one}{\@ne}%
\def\set@footnotewidth{\onecolumngrid}
\def\footnoterule{\kern-6pt\hrule width 1.5in\kern6pt}%
}

\renewcommand\twocolumngrid{
        \def\footnoterule{
        \dimen@\skip\footins\divide\dimen@\thr@@
        \kern-\dimen@\hrule width.5in\kern\dimen@}
        \do@columngrid{mlt}{\tw@}
}%

\makeatother    

\begin{document}
\title{Stripped-Envelope Supernovae for QCD Axion Detection}

\author{Francisco R.\ Cand\'on \orcidlink{0009-0002-3199-9278}} 
\affiliation{Fakult\"at f\"ur Physik, TU Dortmund, Otto-Hahn-Str.~4, 44221 Dortmund, Germany}
\affiliation{CAPA \& Departamento de F\'isica Te\'orica, Universidad de Zaragoza, C. Pedro Cerbuna 12, 50009 Zaragoza, Spain}

\author{Damiano F.\ G.\ Fiorillo \orcidlink{0000-0003-4927-9850}} 
\affiliation{Istituto Nazionale di Fisica Nucleare (INFN), Sezione di Napoli,
Complesso Universitario di Monte Sant’Angelo, Via Cintia, 80126 Napoli, Italy}
\affiliation{Deutsches Elektronen-Synchrotron DESY,
Platanenallee 6, 15738 Zeuthen, Germany}

\author{\'Angel Gil Muyor \orcidlink{0000-0003-0205-3010}}
\affiliation{Dipartimento di Fisica e Astronomia, Universit\`a degli Studi di Padova,\\ Via Marzolo 8, 35131~Padova, Italy}
\affiliation{Istituto Nazionale di Fisica Nucleare (INFN), Sezione di Padova,\\ Via Marzolo 8, 35131~Padova, Italy}

\author{Hans-Thomas Janka \orcidlink{0000-0002-0831-3330}}
\affiliation{Max-Planck Institut f\"ur Astrophysik, Karl-Schwarzschild-Str.~1, 85748 Garching, Germany}

\author{\\ Georg G.\ Raffelt \orcidlink{0000-0002-0199-9560}}
\affiliation{Max-Planck-Institut f\"ur Physik, Boltzmannstr.~8, 85748 Garching, Germany}

\author{Edoardo Vitagliano \orcidlink{0000-0001-7847-1281}}
\affiliation{Dipartimento di Fisica e Astronomia, Universit\`a degli Studi di Padova,\\ Via Marzolo 8, 35131~Padova, Italy}
\affiliation{Istituto Nazionale di Fisica Nucleare (INFN), Sezione di Padova,\\ Via Marzolo 8, 35131~Padova, Italy}


\begin{abstract}
QCD axions would be copiously produced in the proto-neutron star formed in a core-collapse supernova (SN). After escaping, they would convert into gamma rays in the Galactic magnetic field and, as recently shown, in that of the progenitor star itself. Here, we show that Type Ibc SNe---whose progenitors  have lost their hydrogen or even helium envelopes---are the optimal targets for this search. The stripped progenitors are much more compact, and they show larger magnetic fields than both red and blue supergiants, the progenitors of Type~IIP/L SNe. If the next galactic SN is of Type~Ibc, \textit{Fermi}-LAT or a similar gamma-ray satellite might be able to discover the QCD axion down to masses as small as $m_a\simeq 10^{-4}\,\rm eV$ (Peccei-Quinn scale $f_a\simeq 10^{11} \,\rm GeV$).
\end{abstract}

\maketitle

\textbf{\textit{Introduction}}---The QCD axion~\cite{Weinberg:1977ma, Wilczek:1977pj, Kim:1979if, Shifman:1979if, Dine:1981rt, Zhitnitsky:1980tq, DiLuzio:2020wdo} is a natural consequence of the Peccei--Quinn solution~\cite{Peccei:1977hh, Peccei:1977ur} to the strong--CP problem (``why is the neutron electric dipole moment so small?''). It might also constitute dark matter, as it can emerge with the required abundance through various mechanisms in the early Universe~\cite{Preskill:1982cy,Abbott:1982af,Dine:1982ah, OHare:2024nmr}. Barring model-dependent cancellations (e.g.~Ref.~\cite{DiLuzio:2017ogq}), the axion famously features derivative couplings to quarks and a two-photon interaction $-\frac{1}{4}g_{a\gamma}a F_{\mu\nu}\tilde{F}^{\mu\nu}=g_{a\gamma} a \bE\cdot \bB$, with $g_{a\gamma}=C_{a\gamma}\alpha /(2\pi f_a)$, where $\alpha$ is the fine-structure constant, $C_{a\gamma}$ a model-dependent numerical factor, and $f_a$ the axion decay constant, related to its mass through $m_{a}\simeq 5.70\,{\rm \mu eV}\, (10^{12}\,{\rm GeV}/f_a)$~\cite{GrillidiCortona:2015jxo, Gorghetto:2018ocs}. The coupling to nucleons implies $m_a\alt10$--20~meV from the cooling of SN~1987A and isolated neutron stars \cite{Lella:2023bfb, Buschmann:2021juv, Caputo:2024oqc, Carenza:2024ehj,Fiorillo:2025gnd}, while the coupling to photons is crucial for dark matter and solar axion searches through their conversion into photons in external magnetic fields~\cite{Sikivie:1983ip, Raffelt:1987im, IAXO:2019mpb, Irastorza:2018dyq, Ringwald:2023yni, Berlin:2024pzi}.

\begin{figure}[h!]
    \centering
    \includegraphics[width=\linewidth]{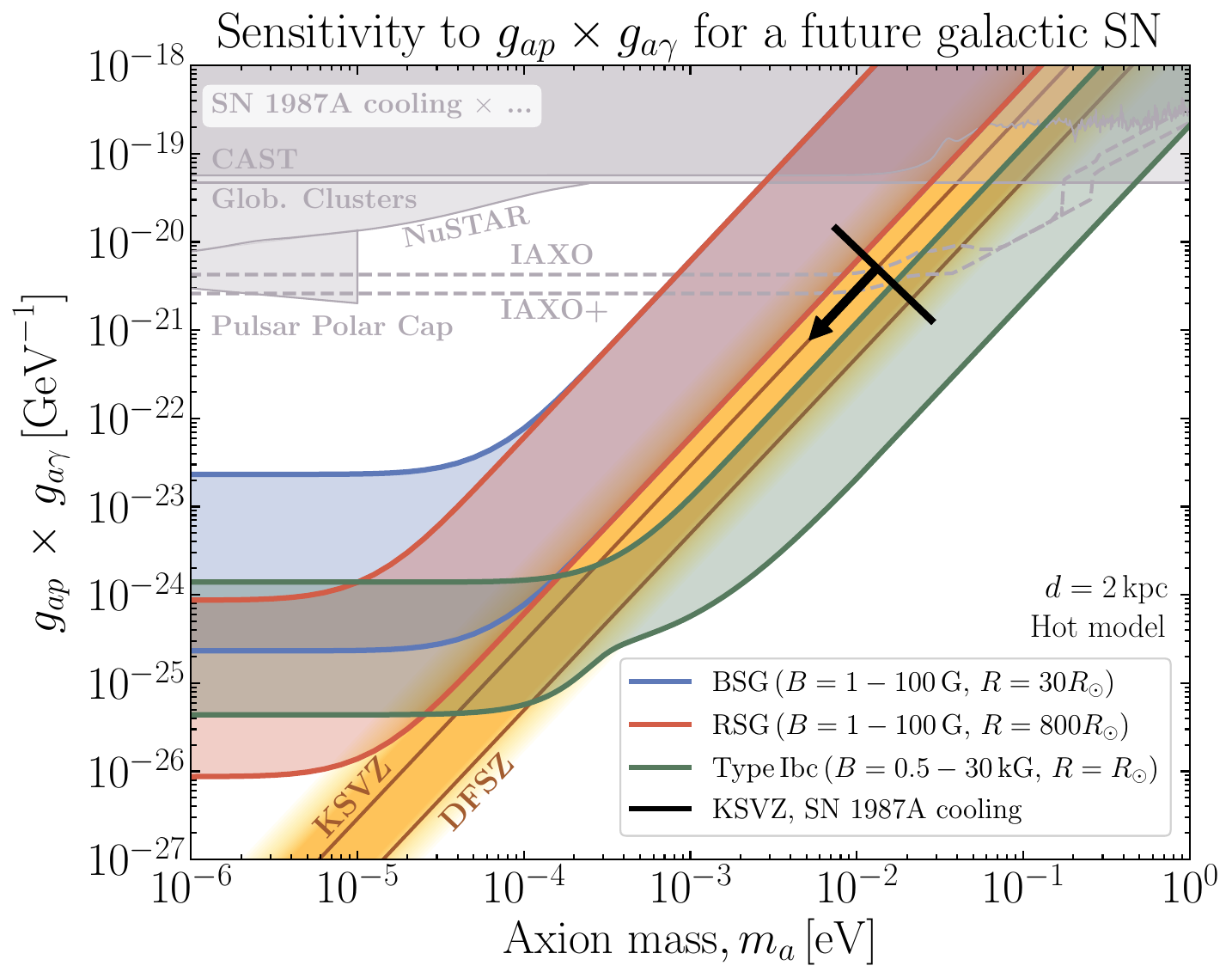}
    \caption{Projected reach on $g_{a p}\times g_{a\gamma}$ for KSVZ axions as a function of $m_a$, assuming detection of 100 MeV $\gamma$ rays from a future galactic SN ($d=2$~kpc) with a satellite like \textit{Fermi}-LAT. The sensitivity bands are for RSG, BSG and stripped-envelope progenitors, assuming the shown values for radius $R$ and dipole field $B$. We assume our hot one-zone SN model for axion production. The black upper bound is the SN~1987A cooling~limit. The gray-shaded limits on $g_{ap}\times g_{a\gamma}$ use this bound of $g_{ap}< 10^{-9}$ together with limits on $g_{a\gamma}$ taken from the compilation of Ref.~\cite{OHare}, specifically bounds from pulsar polar-cap cascades~\cite{Noordhuis:2022ljw}, NuSTAR~\cite{Ruz:2024gkl}, globular clusters~\cite{Dolan:2022kul}, and CAST~\cite{CAST:2024eil}. We also include the projected sensitivities for IAXO and IAXO+ \cite{IAXO:2019mpb} for fixed $g_{ap}=10^{-9}$.
    The QCD axion band is shown in yellow.  
        }
    \label{fig:HotBounds}
    \vskip-6pt
\end{figure}

Likewise, supernova (SN) axions can convert into detectable 100-MeV range photons in astrophysical $B$ fields. The SMM $\gamma$-ray satellite, active during SN~1987A, did not see excess counts at the time of neutrino detection, constraining axion-like particles~\cite{Brockway:1996yr, Grifols:1996id, Payez:2014xsa, Hoof:2022xbe, Manzari:2024jns, Fiorillo:2025gnd}, but its effective area was too small, and SN~1987A, in the Large Magellanic Cloud, was too far away to detect QCD axions. Today, the \textit{Fermi} Large Area Telescope (LAT) achieves an effective area roughly 100 times larger~\cite{Fermi-LAT:2009ihh}, and the next SN could occur much closer, potentially providing a realistic detection opportunity.

It was recently recognized~\cite{Manzari:2024jns} that the $B$ field of the SN progenitor itself, which remains unaffected by the core collapse for several hours, could suffice, in contrast with the Galactic field. Besides the larger $B$ field, the shorter scale of spatial variation is key, facilitating conversion with larger $a$--$\gamma$ momentum mismatch. Therefore, larger $m_a$ values become reachable, shifting sensitivity toward the QCD axion band (Fig.~\ref{fig:HotBounds}), a parameter range notoriously difficult to reach with any method.

In this Letter, we identify Type Ibc SNe as the most promising targets for this approach. Their progenitors had their hydrogen (Type Ib) or even helium (Type Ic) envelopes shed or stripped by a companion. The resulting stars are much more compact, 0.1--$1\,R_\odot$ \cite{Tauris:2015xra}, than either blue or red supergiants (BSGs or RSGs), the progenitors of the more common Type~IIP/L SNe, which have radii of tens or hundreds of $R_\odot$, respectively.
Likewise, stripped-envelope SNe are also ideal targets for detecting decay products of heavy non-standard particles~\cite{Candon:2025ypl}.

The smaller size of the stripped-envelope progenitors is accompanied by larger surface $B$ fields, even though the uncertainties are large for any class of progenitors, as we will discuss. Still, for Type Ibc progenitors, which comprise up to over a third of all core-collapse events---much more frequent in massive galaxies and the local Universe than CCSNe from BSGs~\cite{Kleiser:2011ms,Pessi:2025wht,Graur+2017,Ma+2025}---the observed $B$ fields are much larger than for Type IIP/L progenitors. Our conclusion is that stripped-envelope systems are the most promising targets for this type of axion search, providing additional motivation for the proposed GALAXIS network of $\gamma$-ray satellites~\cite{Manzari:2024jns}. 

Throughout this Letter we use rationalized natural units with $\hbar=c=k_B=1$, where the fine-structure constant is $\alpha=e^2/(4\pi)\simeq1/137$.

\textbf{\textit{Axion production in the SN core}}---To evaluate the reach of such a QCD axion search, we follow the pipeline from production in the nuclear-density SN core, conversion into $\gamma$ rays in the progenitor's $B$ field, to detection in a $\gamma$-ray satellite with \textit{Fermi}-LAT characteristics.  Our main goal is to estimate a realistic detection threshold, which we define by the probability of registering at least three 100 MeV range photons within a few seconds after collapse. In contemporary detectors, the neutrino signal will consist of a detailed time profile over many seconds \cite{Mirizzi:2015eza, SNEWS:2020tbu, Raffelt:2025wty}, defining our observation window, in which the $\gamma$ signal is practically background free.

The next nearby SN will be observed in unprecedented detail in the electromagnetic, neutrino, and probably gravitational-wave channels, as well as nucleosynthesis yields, besides a possible axion signal from deep within the core, testing current SN theory in extraordinary detail. On the other hand, to forecast a marginal axion signal, we have recently argued that it is enough to consider the most salient average features of the source, depending on generic characteristics---essentially, if it is ``hot'' or ``cold'' within a plausible range of final neutron-star masses and nuclear equations of state~\cite{Fiorillo:2025gnd}. Table~\ref{tab:one_zone} shows the average properties of two extreme cases, where the lapse factor $(1+z)^{-1}$ is the average gravitational redshift, and $Mt$ is the nominal mass exposure that multiplies the average axion number emission rate per unit time and unit mass to obtain the total number of emitted axions, $N_a$.

At energies far below $f_a$, QCD axions generically couple to nucleons through the derivative interaction structure $\mathcal{L}_{aNN}=(g_{aN}/2m_N)\,\overline{N}\gamma^\mu \gamma^5 N \,\partial_\mu a$. As a reference case, we use hadronic axions without direct couplings to quarks or leptons, and specifically, the KSVZ model, where the interaction with neutrons nearly vanishes, whereas the one to protons is $g_{ap}=-0.47\,(m_p/f_a)$ \cite{GrillidiCortona:2015jxo}, equivalent to $g_{ap}=-0.77\times10^{-10}\,(m_a/{\rm meV})$. We ignore in-medium modifications~\cite{Springmann:2024mjp} in view of large other uncertainties. The coefficient for the axion-photon coupling is $C_{a\gamma}=-1.92$ so that $g_{ap}\times g_{a\gamma}=3.0\times10^{-17}~{\rm GeV}^{-1}\,(m_a/{\rm eV})^2$, defining the KSVZ line in Fig.~\ref{fig:HotBounds}. Using $m_a<13$~meV as a nominal SN~1987A cooling bound \cite{Raffelt:2006cw, Carenza:2019pxu, Lella:2023bfb} yields $g_{ap}<1.0\times10^{-9}$ and $g_{ap}\times g_{a\gamma}<5\times10^{-21}~{\rm GeV}^{-1}$, shown in Fig.~\ref{fig:HotBounds} by the black line and arrow.

\begin{table}
\vskip-6pt
\caption{Physical parameters of our cold and hot average one-zone SN models~\cite{Fiorillo:2025gnd}, the corresponding total number of emitted axions, $N_a$, and their average energy, $E_{\rm av}$.}
\begin{tabular*}{\columnwidth}{@{\extracolsep{\fill}}lll@{}}
\toprule
\textbf{Quantity} & \textbf{Cold} & \textbf{Hot}\\
\midrule
Density $\rho$ [$10^{14}~\mathrm{g/cm^3}$] & 4.0 & 6.0 \\
Temperature $T$ [MeV] & 30 & 45\\
Proton fraction $Y_p$ & 0.15 & 0.15\\
Lapse $(1+z)^{-1}$ & 0.75 & 0.65\\
Exposure of mass $M t$ [$M_\odot\,\mathrm{s}$] & 5.0 & 10.0\\
\midrule
\multicolumn{3}{l}{Bremsstrahlung emission parameters}\\
\quad $N_a$ [$g_{ap}^2 10^{73}$] & 34.9 & 235\\
\quad $E_{\rm av}$ [MeV]& 55.7 & 74.3\\
\quad Pinching parameter $\alpha$ & 1.92 & 2.06\\
\bottomrule
\end{tabular*}
\label{tab:one_zone}
\end{table}

We use $NN\to NNa$ bremsstrahlung as a source, based on a parametric rate prescription detailed earlier~\cite{Fiorillo:2025gnd}. The result is surprisingly similar to more involved calculations, which, however, do not guarantee greater reliability due to uncertain nuclear-physics inputs.
Our time-integrated axion flux is well approximated by a quasi-thermal distribution of the form
\begin{equation}
    \frac{dN_a}{dE_a}=N_a\, \frac{(\alpha+1)^{\alpha+1}}{\Gamma(\alpha+1)}\,
    \frac{E_a^\alpha}{\Eav^{\alpha+1}}\,
    \exp\left[-\frac{(1+\alpha)E_a}{\Eav}\right],
\end{equation}
which is normalized to $N_a$, has the average energy $\langle E_a\rangle=\Eav$, and uses a ``pinching parameter'' that would be $\alpha=2$ for a Maxwell-Boltzmann distribution. We provide our fit parameters for the cold and hot models~\cite{Fiorillo:2025gnd} in Table~\ref{tab:one_zone} as a function of $g_{ap}$. The spectral shape differs between bremsstrahlung prescriptions, but it is difficult to assess in detail because the required dynamical spin and isospin structure functions of the nuclear medium are poorly known. However, to estimate the margin of axion detection, based on a few detected $\gamma$ rays, does not depend on spectral details.

We do not include pionic axion production, $\pi^- p \to n a$, which has been proposed as a mechanism for enhanced axion emission in recent years \cite{Carenza:2020cis, Fischer:2021jfm, Lella:2023bfb}. However, the measured pion-nucleon phase shifts imply an upward shift of the pion energy exceeding 100~MeV \cite{Fore:2023gwv}---consistent with simple textbook estimates~\cite{Ericson:1988gk}---and seem to disfavor a large thermal pion population \cite{Fiorillo:2025gnd}. If axion emission were nevertheless enhanced, it would only increase the prospects for QCD axion detection.

\textbf{\textit{Axion--photon conversion around stars}}---Axions convert into photons in the magnetic field of the progenitor star, as proposed in Ref.~\cite{Manzari:2024jns}. The resulting axion-photon conversion probability $P_{a\gamma}$ depends sensitively on the field strength and structure. We here briefly summarize our earlier detailed results~\cite{Fiorillo:2025gnd} for the parametric dependence of $P_{a\gamma}$ in an assumed dipole field.

The relevant length scale for axion-photon conversion is the coherence length $\ell$ of the $B$ field; for dipole structure, this is roughly the photosphere radius $R$, at which axion conversion into gamma rays can proceed unimpeded by plasma refraction~\cite{Fiorillo:2025gnd}. The other relevant scale is the length over which ultra-relativistic axions and photons remain coherent, characterized by $|\Delta_a|=m_a^2/2E_a$, the momentum mismatch at a given energy.

If $|\Delta_a| R\ll 1$, the conversion probability is energy independent, $P_{a\gamma}\simeq g_{a\gamma}^2 B^2 R^2/16$, where $B$ is the field strength at the photosphere $R$. The conversion accumulates over distances comparable with $R$, so this expression depends on the assumed dipole structure only by a numerical factor. In the opposite limit of $|\Delta_a| R\gg 1$, axion conversion is a local process, since coherence is not preserved over distances larger than $|\Delta_a|^{-1}$, and
\begin{equation}\label{eq:conversion_probability}
    P_{a\gamma}\simeq g_{a\gamma}^2 B^2 E_a^2/m_a^4.
\end{equation}
These expressions capture the parametric dependence; for our detailed calculations, we use numerical solutions of \hbox{$a$--$\gamma$} conversion in a dipole field, assumed orthogonal to the line of sight. We also include non-linear QED refraction in the strong $B$ field, which for Ibc progenitors can be so intense as to affect the conversion probability, as seen in the dent of the lower green curve in Fig.~\ref{fig:HotBounds}.

To probe the QCD axion, the most interesting regime is the massive one, where $|\Delta_a|R\gg 1$. Due to the characteristic $m_a$ dependence of the conversion probability, the sensitivity band is parallel to the axion line (Fig.~\ref{fig:HotBounds}). Axion-photon conversion is now local, and thus, depends only on the surface magnetic field.

\textbf{\textit{Observational status of magnetic fields}}---Strip\-ped-envelope SN progenitors are generally thought to be Wolf-Rayet (WR) stars---massive stars that have lost their hydrogen (or even helium) envelopes through stellar winds or interactions in binary systems. Wind broadening of emission lines is the main challenge to reliable $B$-field detection. For 11~WR stars, Ref.~\cite{2014ApJ...781...73D} finds marginal detection of hundreds of G in three of them, and upper bounds of 500~G for the others. Refs.~\cite{2016MNRAS.458.3381H,2020MNRAS.499L.116H,2023MNRAS.524L..21J} also measure $B$ fields of hundreds of G, but explicitly warn that these pertain only to the regions where the lines are formed, i.e., significantly outside of the progenitor. Hence, the surface field is likely much larger. In 2023, a stunning value of 43~kG was reported from the surface of HD~45166~\cite{2023Sci...381..761S}. It thus appears likely that WR stars may exhibit coherent surface $B$ fields from hundreds of G to tens~of~kG.

Comparing with RSGs, the most common SN progenitors, the few reported fields are all in the range 1--10\,G (Ref.~\cite{2017A&A...603A.129T} and references); historically, the first case was Betelgeuse~\cite{2010A&A...516L...2A}. Dorch~\cite{Dorch:2004af} has proposed that these fields may originate from the saturation of a dynamo mechanism to super-equipartition in the convective mantle. Surface patches could then even reach up to 500\,G, but the filling fraction is small. The field is highly intermittent, with a large fraction of the surface having 50\,G, and perhaps 10\% above 100\,G. It appears fair to say that RSGs tend to have weak coherent $B$ fields, with local small-scale values perhaps reaching up to 100\,G.

Finally, BSGs---perhaps only 1\% of all core-collapse progenitors~\cite{Graur+2017, Ma+2025}---were the main targets proposed in Ref.~\cite{Manzari:2024jns}. However, the quoted 100\,G--1\,kG fields seem overly optimistic because, while Refs.~\cite{2015A&A...574A..20F,2009ARA&A..47..333D,2015A&A...584A..54P} do mention such strong fields, these pertain to massive main-sequence stars. The absence of such large field detections in BSGs is attributed to their much larger size, which by flux conservation suggests much smaller fields~\cite{2015A&A...584A..54P,2017MNRAS.465.2432G}. A positive detection exists for $\zeta$~Ori~A~\cite{2008MNRAS.389...75B}, for which observations of 50--100~G are quoted in the introduction of the MiMes survey \cite{2017MNRAS.465.2432G}. Overall, observations suggest rather weak BSG fields of tens of G, exceptionally perhaps up to 100\,G.

All these types of stars are surrounded by winds, which can strongly affect the $B$-fields and, specifically, modify the assumed dipole structure. Stellar winds thus set an upper radius for which the dipole structure holds, a question discussed in the End Matter.

\textbf{\textit{Prospects for QCD axion discovery}}---Figure~\ref{fig:HotBounds} summarizes our detection forecast for stripped-envelope SNe, compared with RSG and BSG cases. The $B$ fields are taken in the ranges discussed above: 1--100\,G for RSG and BSG, and 500\,G--30\,kG for Ibc SNe. We determine the $\gamma$-ray fluence at Earth by multiplying the axion spectrum from the hot SN model with the conversion probability from a numerical integration as in Ref.~\cite{Fiorillo:2025gnd}. We finally obtain the \textit{Fermi}-LAT event counts, using the on-axis effective area extracted from the \textit{Fermi} Science Tools (\textsc{\texttt{fermitools}}) for the \textsf{\texttt{P8R3\_TRANSIENT020\_V3}} class, shown in the End Matter. We set a detection threshold when the energy-integrated number of expected counts exceeds $N_\gamma>3$, a round number close to the Feldman-Cousins upper bound on the expected event counts in the case of no observation~\cite{Feldman:1997qc}. In the End Matter, we complement our detection forecast with existing bounds for axion dark matter together with sensitivities of future haloscope searches.

For our chosen distance of 2~kpc, the detection prospects with SNe Ibc look quite promising, reaching all the way to $m_a\agt0.1$--1\,meV, whereas RSGs and BSGs, even with our most optimistic parameters, do not touch the KSVZ line. For BSGs and RSGs, our more pessimistic forecast compared with Ref.~\cite{Manzari:2024jns} derives from our smaller assumed $B$ fields and the absence of pionic emission. On the other hand, the
number $N_a$ of emitted axions from the hot SN model is optimistic; the cold model reduces $N_a$ by a factor of 7, yielding a $\sqrt{7}$ worse $g_{ap}\times g_{a\gamma}$ sensitivity (see End Matter). SNe~Ibc still cover the axion band.

The detection prospects essentially hinge on the diagonal bands in Fig.~\ref{fig:HotBounds}, the region of large $m_a$, where the conversion probability is given by Eq.~\eqref{eq:conversion_probability}. Multiplying with the axion spectrum and the \textit{Fermi}-LAT effective area yields the number of event counts as
\begin{equation}
    N_\gamma=3.24\times 10^{41}\, {\rm GeV}^2\ \frac{g_{ap}^2 g_{a\gamma}^2 B_{\mathrm{G}}^2}{m_{\mathrm{meV}}^4 d_{\rm kpc}^2},
\end{equation}
where $m_{\mathrm{meV}}=m_a/1\,\mathrm{meV}$, $B_{\mathrm{G}}=B/1\,\mathrm{G}$, and $d_{\rm kpc}$ is the SN distance in kpc. Since for the KSVZ axion $g_{ap} g_{a\gamma}=3.0\times10^{-23}\, {\rm GeV}^{-1}m_\mathrm{meV}^2$, the detection threshold is encoded in the single parameter $B_{\mathrm{G}}/d_{\mathrm{kpc}}\gtrsim103$, depending only on local $B$-field properties. The only condition is that the field strength is achieved outside of the photosphere, and is maintained for a coherence length $\ell\gtrsim \Delta_a^{-1}\simeq 0.06\, R_\odot\, E_{100}\, m_{\mathrm{meV}}^{-2}$, where $E_{100}$ is $E_a/100\,{\rm MeV}$.

\textbf{\textit{Observed signal and mass reconstruction}}---It is unlikely that a possible axion signal will be precisely near the detection threshold; one may actually observe enough events to reconstruct a spectrum. As an example, we define a benchmark galactic SN Ibc located at the coordinates of the WR star HD~45166 ($ \text{RA} = 96^\circ\,34'\,47.3'', \; \text{Dec} = +07^\circ\,58'\,28.1'' $) at a distance of 2~kpc and a surface magnetic field of 30~kG. The mock data were simulated with the \textsc{\texttt{fermitools}} (v.~\textsc{\texttt{2.4.0}}) using the observation simulation tool \textsf{\texttt{gtobssim}} \cite{fermitools}. We selected Pass~8 photon events with the \textsf{\texttt{P8R3\_TRANSIENT020\_V3}}
Instrument Response Functions, event class 16, event type 3, a zenith-angle cut of $z_{\max}=100^\circ$, and the quality filter \textsf{\texttt{(DATA\_QUAL>0)\&\&(LAT\_CONFIG==1)}}, in the energy range from $20~\mathrm{MeV}$ to $3~\mathrm{GeV}$.

\textit{Fermi}-LAT operates in survey mode, meaning that it continuously scans the entire sky. At any given time, roughly 20\% of the sky lies within its field of view (FoV). Consequently, the instrument’s effective area depends not only on energy, but also on incidence angle relative to the telescope’s axis. To be specific, we selected an observation time when our source was within the LAT field of view, at an average off-axis angle of $\langle \theta \rangle \sim 52^{\circ}$ (2024~Mar~8, 06:12:00~UTC), which corresponds to a reduction of approximately $50\%$ in the effective area with respect to the on-axis values used for Fig.~\ref{fig:HotBounds}.

\begin{figure}
    \centering
    \includegraphics[width=\linewidth]{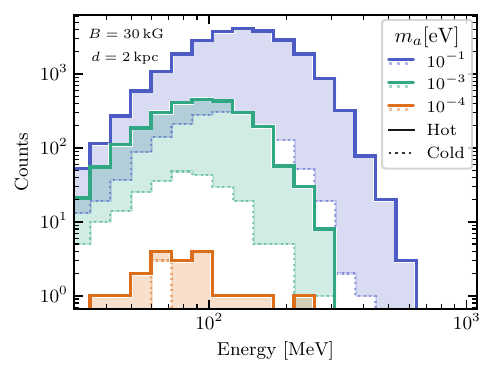}
    \caption{Simulated \textit{Fermi}-LAT spectra expected for KSVZ axions with different $m_a$ values, using both the hot model (thick histograms) and the cold model (dotted histograms). We used the celestial coordinates of HD~45166 at a period with an average off-axis angle of $\langle \theta \rangle \sim 52^{\circ}$. The energy bins match a resolution of 20\% of the photon energy.}
    \label{fig:SimulSpectra}
\end{figure}

Figure~\ref{fig:SimulSpectra} shows the event spectrum for a KSVZ axion with varying $m_a$. Masses near the detection threshold produce a small count rate, which peaks below 100\,MeV, even for the hot SN core. On the other hand, large $m_a$ lead to a much larger count rate, peaking closer to 200\,MeV, due to the rapidly increasing conversion probability $P_{a\gamma}\propto E_a^2$. Hence, while purely speculative since we do not know the properties of the next galactic SN, a high-statistics signal might provide hints on~$m_a$.

\textbf{\textit{Discussion and outlook}}---``Invisible axions'' solving the strong-CP problem are notoriously difficult to detect, the main hope resting on direct dark matter searches. Stellar cooling arguments require $m_a\alt10$~meV, whereas black-hole superradiance suggests $m_a\agt10$~peV~\cite{Caputo:2024oqc}. Near the high-mass end of this allowed range, sometimes called the meV frontier, a solar axion search used to be the only approach---independent of the dark matter assumption---with a realistic detection perspective.

The proposal to search for gamma rays coincident with the neutrino burst from the next Galactic SN is a novel approach, motivated by the possibility of axion-photon conversion in the progenitor’s magnetic field. This strategy requires a coordinated network of gamma-ray satellites providing continuous, full-sky coverage. Such a system would entail substantial effort and investment, but presumably would also serve a broad range of other astrophysical objectives and targets.

Waiting for the next Galactic SN may take a long time, but constructing IAXO for solar axion detection is itself a decades-long endeavor, and dark matter searches are similarly career spanning. Therefore, any new approach merits serious consideration and discussion, even if it requires taking a long-term perspective.

The magnetic fields of SN progenitors are among the least understood elements in the production–detection pipeline of this method, motivating dedicated studies. We identify here a previously overlooked class of targets: stripped-envelope stars, the progenitors of Type Ibc SNe, which comprise up to a third of all core-collapse events~\cite{Pessi:2025wht}. Their smaller geometric size and correspondingly larger $B$ fields make them stand out among conceivable targets. Blue supergiants are much rarer, perhaps 1\% of all cases~\cite{Kleiser:2011ms,Pessi:2025wht}, whereas the more familiar red supergiants tend to have weaker $B$ fields, and therefore, must be closer for a realistic QCD axion detection.

Assuming nuclear bremsstrahlung as the only production process, the discovery potential is summarized in a single number---the ratio of the progenitor magnetic field $B$ over the SN distance $d_{\rm SN}$. A QCD axion detection in the form of a few 100 MeV gamma rays in a satellite similar to \textit{Fermi}-LAT can be achieved for $B/{\rm kG}\agt d_{\rm SN}/10~{\rm kpc}$. Therefore, a Type~Ibc SN at a Galactic-center distance would require a kG-range $B$ field. For stripped-envelope stars, this possibility is real---rare cases even go beyond this limit. The compact nature of stripped-envelope progenitors allows one to touch $m_a$ down to $10^{-4}\,\mathrm{eV}$.

The next Galactic SN was always expected to become a treasure trove of scientific discoveries, with the potential detection of QCD axions among its richest rewards.

\textbf{\textit{Acknowledgements---}}This Letter is
based upon work from COST Action COSMIC WISPers
(CA21106), supported by COST (European Cooperation
in Science and Technology). DFGF is supported by the Alexander von Humboldt Foundation (Germany). In Munich, we acknowledge partial support by the German Research Foundation (DFG) through the Collaborative Research Centre ``Neutrinos and Dark Matter in Astro- and Particle Physics (NDM)'', Grant SFB-1258-283604770, and under Germany’s Excellence Strategy through the Cluster of Excellence ORIGINS EXC-2094-390783311. In Padua, we acknowledge support by the Italian MUR Departments of Excellence grant 2023--2027 ``Quantum Frontier'' and by Istituto Nazionale di Fisica Nucleare (INFN) through the Theoretical Astroparticle Physics (TAsP) project. AGM acknowledges the support of FCT -- Fundação para a
Ciência e Tecnologia, I.P., with DOI identifiers 10.54499/2023.11681.PEX, and the project 10.54499/2024.00249.CERN.  EV acknowledges support from the Italian Ministero dell'Università e della Ricerca through the FIS 2 project FIS-2023-01577 (DD n.~23314 10-12-2024, CUP C53C24001460001).

\bibliography{refs}
\bibliographystyle{bibi}

\onecolumngrid
\newpage
{\ }

\begin{center}
\textbf{\large End Matter}
\end{center}

\twocolumngrid

\textbf{\textit{Fermi-LAT response functions}}---We here report the on-axis effective area adopted in the main text. It is extracted from the \textsf{\texttt{8R3\_TRANSIENT020\_V3}} calibration files included in \textsc{\texttt{fermitools}} v2.4.0 \cite{fermitools}, which correspond to the most up-to-date Instrument Response Functions for this event class. The smallest incidence angle available in these FITS files is $\theta = 9^{\circ}$, which we consider as representative of the most ideal (near on-axis) observational configuration. 

\begin{figure}[h!]
    \centering
    \includegraphics[width=\linewidth]{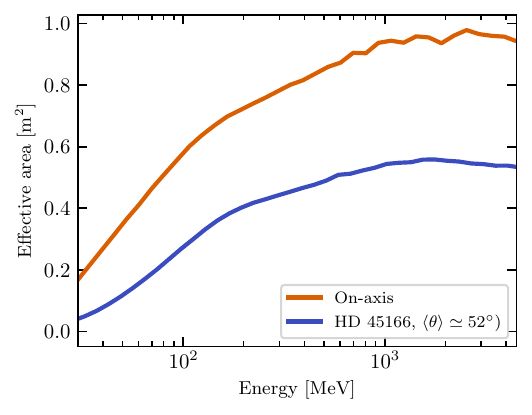}
    \caption{Comparison between \textit{Fermi}-LAT's \textit{on-axis} effective area extracted from the \textsf{\texttt{P8R3\_TRANSIENT020\_V3}} Instrument Response Functions and the average effective area obtained with \textsf{\texttt{gtexposure}} for our simulated observation at $\langle \theta \rangle \sim 52^{\circ}$.}
    \label{fig:Aeff}
\end{figure}

For comparison, we use the \textsf{\texttt{gtexposure}} tool from the \textsc{\texttt{fermitools}} suite to compute the effective area corresponding to our simulated dataset in Fig.~\ref{fig:SimulSpectra}, which yields an average inclination angle of $\langle \theta \rangle \sim 52^{\circ}$. Both effective areas are shown in Fig.~\ref{fig:Aeff}. 
Our choice of a larger inclination angle is mostly for illustration, as it shows the impact that the SN position in the sky can have on the LAT sensitivity. 
Of course, the worst scenario occurs when the SN event lies outside the LAT field of view (FoV), which covers approximately 20\% of the sky at any given instant. An instrument like the proposed GALAXIS~\cite{Manzari:2024jns}, would circumvent this problem with a $4\pi$ FoV. Furthermore, if the explosion happens while the LAT is traversing the South Atlantic Anomaly, the corresponding data would fall outside the Good Time Intervals.

\begin{figure}
    \centering
    \includegraphics[width=\linewidth]{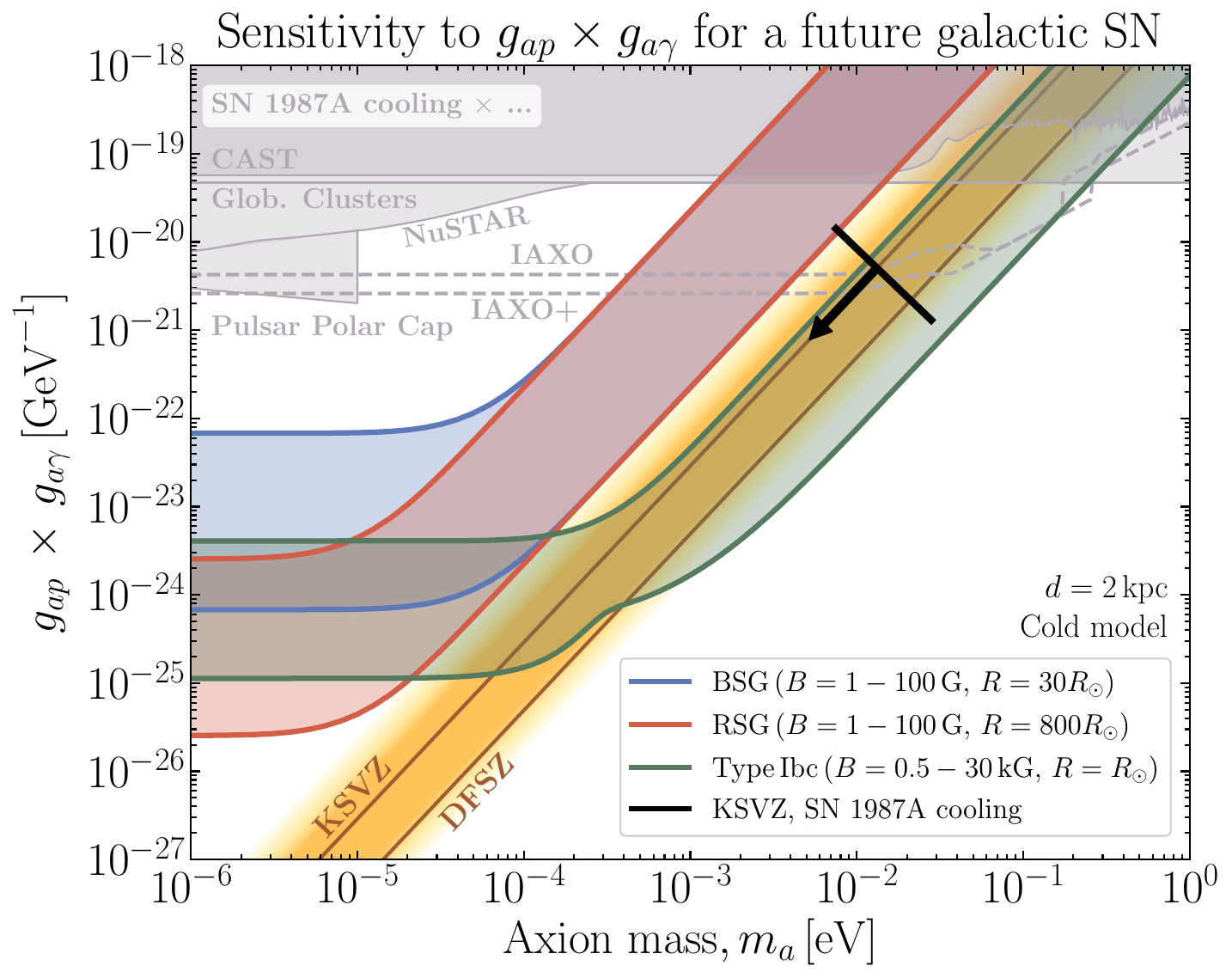}
    \caption{Like Fig.~\ref{fig:HotBounds}, but assuming instead our cold SN model (see Table~\ref{tab:one_zone}) for the production of axions from the core.}
    \label{fig:ColdBounds}
\end{figure}
\textbf{\textit{Impact of the SN model}}---For the results shown in Fig.~\ref{fig:HotBounds}, we chose our hot SN model, specified in Table~\ref{tab:one_zone}, near the upper end of the plausible emitted total axion number $N_a$ for bremsstrahlung-only production. On the other extreme is the cold model, with a factor of 7 fewer axions, worsening the sensitivity on $g_{ap}\times g_{a\gamma}$ by $\sqrt{7}$. This simple estimate is confirmed by Fig.~\ref{fig:ColdBounds}, where we show the projected reach for the cold model, all else unchanged.

\begin{figure}[ht]
    \centering
    \includegraphics[width=\linewidth]{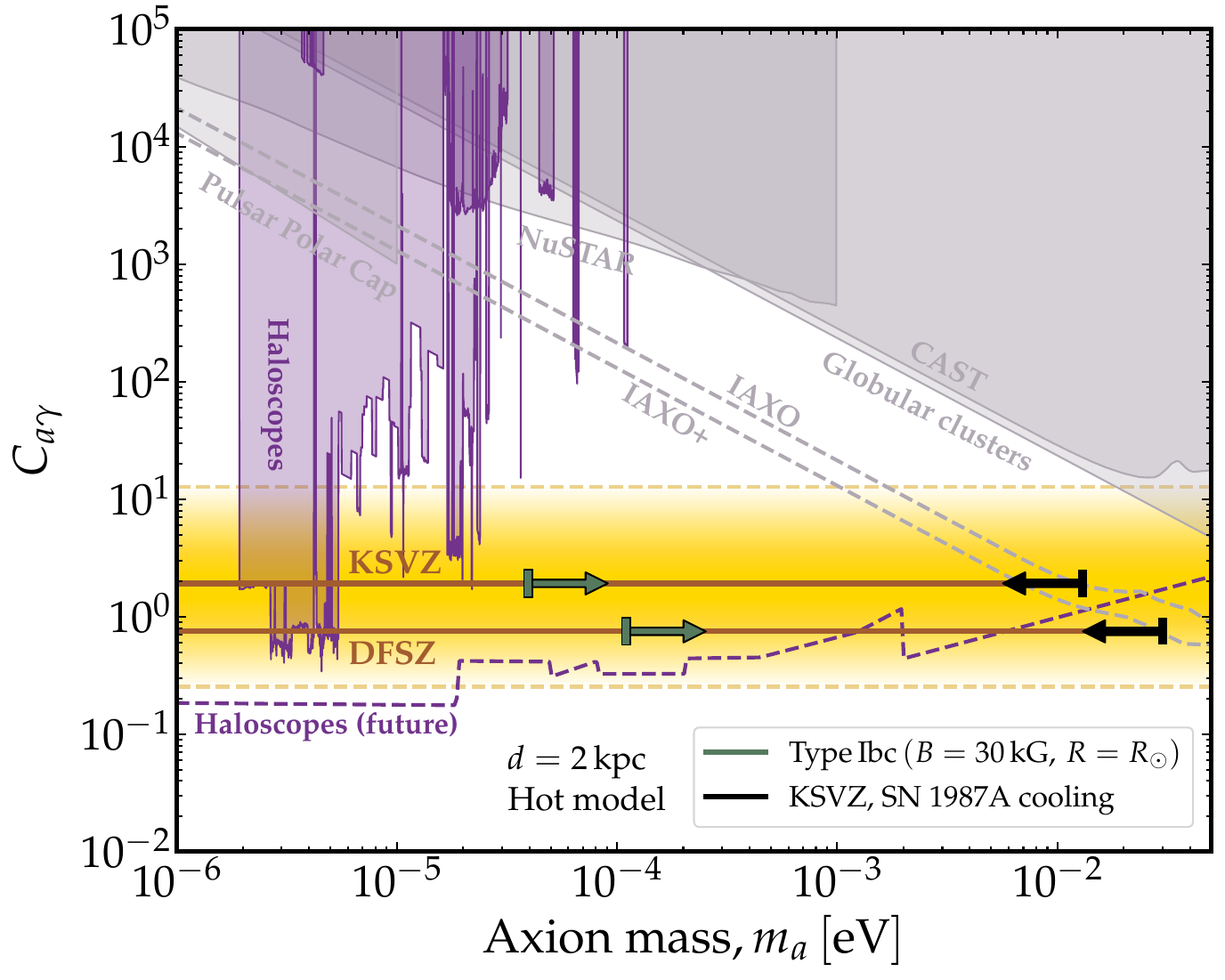}
    \caption{Projected reach on $C_{a \gamma}$ for KSVZ axions, and for DFSZ axions with no coupling to neutrons, assuming detection of 100 MeV $\gamma$ rays from our most optimistic case of a stripped-envelope SN. The sensitivity range is denoted by the arrows on the KSVZ and DFSZ lines, where the black ones represent the SN~1987A cooling~limit. The gray-shaded regions are limits from other sources, while the purple ones assume that axions are the Galactic dark matter. We show bounds from pulsar polar-cap cascades~\cite{Noordhuis:2022ljw}, NuSTAR solar observations~\cite{Ruz:2024gkl}, globular clusters~\cite{Dolan:2022kul}, and CAST~\cite{CAST:2024eil}. We also show  the future IAXO and IAXO+ sensitivities \cite{IAXO:2019mpb}. The dark matter limits include conversion around neutron stars \cite{Foster_2022, Battye_2023}, experimental searches \cite{ADMX:2025vom,Bae:2024kmy,Adair:2022rtw,Hoshino:2025fiz,Grenet:2021vbb,HAYSTAC:2024jch,Garcia:2024xzc,Quiskamp:2024oet,QUAX:2024fut,Ahyoune:2024klt,Wuensch:1989sa,TASEH:2022vvu,Hagmann:1996qd}, and claimed future sensitivities \cite{Graham:2015ouw,Millar:2022peq,Ahyoune:2023gfw,BRASS,Liu:2021pei,Aja:2022csb,DeMiguel:2023nmz,Fan:2024mhm,Beurthey:2020yuq,Alesini:2023qed,QUAX,McAllister:2017lkb}. 
    }
    \label{fig:BoundsDM}
\end{figure}

\textbf{\textit{Comparison with axion dark matter searches}}---The production, conversion, and detection mechanisms described here do not depend on the QCD axion being dark matter. However, the detection of 100 MeV $\gamma$ rays from a future galactic SN would certainly provide a tantalizing hint for such a role, particularly as $m_a$ would fall in the range predicted in post-inflationary scenarios (see, e.g.,\ Refs.~\cite{Gorghetto:2020qws,Kim:2024wku,Benabou:2024msj,Saikawa:2024bta}). In Fig.~\ref{fig:BoundsDM}, we show our most optimistic sensitivity (a stripped-envelope progenitor at $d=2$~kpc with $B=30$~kG and $R=1\,R_\odot$, assuming our hot one-zone SN model), together with axion dark matter bounds and future experiment sensitivities. Our detection prospects are complementary to future haloscope searches for a range of axion masses of several orders of magnitude.

\textbf{\textit{Impact of stellar winds on \boldmath{$B$}-fields}}---One of the astrophysical uncertainties that could affect the results presented here is related to the presence of stellar winds. This outflow of material from the stellar surface stretches the field lines in the radial direction, breaking the dipole structure as the field enters the Parker regime, so that the radial component becomes $B_{ r}\propto r^{-2}$. Therefore, the wind increases the component parallel to the line of sight at the cost of the transverse component, which is required for the axion-photon conversion.

The impact of the wind is measured by the ratio between the magnetic pressure and the ram pressure, the so-called magnetic confinement parameter, whose value at the stellar surface is~\cite{ud-Doula:2002obk}
\begin{equation}
    \eta=\frac{B^2}{2}\left(\rho_w\frac{v_w^2}{2}\right)^{-1}=\frac{4\pi B^2 R^2}{\dot{M}v_{ w}},
\end{equation}
where $\rho_w=\dot{M}/4\pi R^2 v_w$ is the wind density, with $\dot{M}$ the rate of emitted mass, and $v_w$ the terminal wind velocity (recall that we use rationalized electromagnetic units).
This quantity defines the radius at which the the magnetic pressure is not enough to keep the field lines against the wind pressure, the so-called Alfvén radius \cite{2022A&A...657A..60S},
\begin{equation}
    R_A=R\left[0.29+(\eta+0.25)^{1/4}\right].
\end{equation}
At larger radii, the dipole structure of the $B$-field is expected to break down progressively. For the typical conditions of a WR star, we have $\dot{M}\sim 10^{-5}\,M_\odot/\mathrm{yr}$, $B\sim 1\,\mathrm{kG}$, $R\sim 1\,R_\odot$, and $v_w\sim 1000\,\mathrm{km/s}$, so that $\eta\sim 0.08$ and $R_A\simeq 1.05 R$. Hence, already a few percent of the progenitor radius above the star, the dipolar structure might be affected by the powerful winds. Under extremely conservative assumptions, we can assume the dipole structure to break completely at $R_A$, with the $B$-field becoming purely radial so that the conversion is completely switched off. This would reduce the conversion probability in the massless-axion regime by a factor of $\bigl[\int_R^{R_A}  dr\, B(r)/\int_R^{+\infty} dr\, B(r)\bigr]^2$, i.e.,
\begin{equation}
    \left(\left.\int_R^{R_A}   \frac{dr}{r^3}  \right/\int_R^{+\infty} \frac{dr}{r^3}\right)^2\simeq 0.09^2,
\end{equation}
which means that the flat part of our sensitivities in Figs.~\ref{fig:HotBounds} and~\ref{fig:ColdBounds} would worsen by roughly an order of magnitude. Correspondingly, the transition between the massless and massive axion regimes happens at a higher value of $m_a$, when $\Delta_a\sim 1/(R_A-R)$, instead of $\Delta_a\sim 1/R$. For the conditions assumed above, the wind shifts the transition between the regimes to $m_a$ values that are $1/\sqrt{0.05}\sim 5$ times larger.

Such an extreme estimate for the impact of winds on the magnetic field structure is overly conservative. To begin with, we assumed that the field becomes purely radial at $R_A$, which is of course unrealistic. Further, the highest values of magnetic field we consider for Type Ic SNe reach up to $B\sim 40\,\mathrm{kG}$, for which the magnetic confinement parameter is much larger and the dipole field would extend to distances comparable to the progenitor radius. We are also neglecting the rotation of the star, which twists the field lines, creating a toroidal component $B_\phi\sim \Omega r B_r/v_w\sim (v_{\rm rot}/v_w) (r/R) B_r$, where $v_{\rm rot}=\Omega R$ is the surface rotation field, and $\Omega$ is the angular velocity~\cite{Parker:1958zz}. The toroidal field adds a contribution to axion-photon conversion at larger distances; for $v_{\rm rot}\sim 100\,\mathrm{km/s}$, this contribution can be comparable to the conversion in the dipolar field immediately outside the star. All these arguments point to much larger conversions. 

Most importantly,  the geometry and potential turbulence of the magnetic field have negligible impact on the high-mass region of the parameter space. At large masses, axion conversion happens over a length scale $\ell\sim E_a/m_a^2$, which is much shorter than the progenitor radius. Therefore, for our forecasts to be valid, it suffices that the magnetic field maintains coherence over a distance $\ell$, which is much shorter than the progenitor radius. Even in the most pessimistic case that the dipolar field extends only up to a small fraction of the progenitor radius $\delta r\sim 0.0015 R\sim 10^8\,\mathrm{cm}$, this would still be enough to efficiently convert axions with masses above $m_a\gtrsim \sqrt{E_a/\delta r}\sim 10^{-3}\,\mathrm{eV}$, which is indeed the range in which we claim that QCD axion detection might happen.

\end{document}